\documentclass[12pt]{article}
\usepackage{graphics,graphicx,color,amsmath,array,hhline,fancyhdr}
\usepackage[latin2]{inputenc}
\usepackage[magyar,english]{babel}
\begin{document}

\title{From Quark Gluon Plasma to\\a Perfect Fluid of Quarks and Beyond}
\author{M. Csanád$^{\, 1}$, T. Csörgő$^{\, 2}$, B. Lörstad$^{\, 3}$, M. Nagy$^{\, 1}$ and A. Ster$^{\, 2}$\\
{\footnotesize $^1$Dept. Atomic Phys., ELTE, H-1117 Budapest, Pázmány P. 1/a, Hungary}\\
{\footnotesize $^2$MTA KFKI RMKI, H - 1525 Budapest 114, P.O.Box 49, Hungary}\\
{\footnotesize $^3$Dept. Physics, University of Lund, S - 22362
Lund, Sweden}}

\maketitle

\begin{abstract}
With high energy heavy ion collisions one tries to create a new
forms of matter that is similar to the one present at the birth of
our Universe. Recent development on flow pattern, initial
energy-density and freeze-out temperature shows that most likely
this new form of matter is in a deconfined state, has colored
degrees of freedom and is more fluid-like than gas-like. In
present paper we calculate estimations on the physical properties
of this new-old matter.
\end{abstract}

\begin{quote}
{\small {\it ``We simply do not yet know enough about the physics
of elementary particles to be able to calculate the properties of
such a melange with any confidence. \dots Thus our ignorance of
microscopic physics stands as a veil, obscuring our view of the
very beginning.''}}

S. Weinberg, about the first hundredth of a second~\cite{Weinberg}
\end{quote}

\section{Introduction}

Ultra-relativistic collisions, so called ``Little Bangs'' of
almost fully ionized Au atoms are observed in four major
experiments at the RHIC accelerator at the highest currently
available colliding energies of $\sqrt{s_{NN}} = 200$ GeV. The aim
of these experiments is to create new forms of matter that existed
in Nature a few microseconds after the Big Bang, the creation of
our Universe.

Quantum Chromodynamics (QCD), the theory of quarks and gluons, the
strong force interacting between them and their color degree of
freedom was formulated and established soon after Weinberg's
famous book~\cite{Weinberg} about the early Universe had been
published. Confinement is an important (though mathematically
never proven) property of QCD, its consequence is that quarks are
bound into hadrons in a matter of normal temperature and pressure.

In the early Universe, energy density was many orders of magnitude
higher than today, and at that high energy densities, deconfined
phases of colored matter might have existed. Quark-gluon plasma
(QGP) is such a phase, that might have existed during the first
few microseconds after the Universe came into existence. This type
of matter was searched for at the SPS, and experiments at RHIC are
continuing this effort. Evidence for formation of a hot and dense
medium in gold-gold collisions was found based on a phenomenon
called jet quenching, and confirmed by its disappearance in
deuteron-gold collisions~\cite{Adler:2003ii}.

A consistent picture emerged after the first three years of
running the RHIC experiment: quarks and gluons indeed become
deconfined, but also behave collectively, hence this hot matter
acts like a liquid~\cite{Adcox:2004mh}, not like an ideal gas
theorists had anticipated when defining the term QGP. The
situation is similar to as if prisoners (quarks and gluons
confined in hadrons) have broken out of their cells at nearly the
same time, but they find themselves on the crowded jail-yard
coupled with all the other escapees. This strong coupling is
exactly what happens in a liquid~\cite{Riordan:2006df}.

\subsection{A sign for hydrodynamic behavior: elliptic flow}

Azimuthal asymmetry of single particle spectra measured in
relativistic heavy ion collisions is called elliptic flow ($v_2$).
It is an indication of liquid-like behavior~\cite{Adler:2003kt},
and can be explained by
hydrodynamics~\cite{Csanad:2003qa,Hama:2005dz,Broniowski:2002wp}.
In the hydrodynamic picture it turns out, that elliptic flow can
result from the initial spatial asymmetry but also from
momentum-space asymmetry. Important is, that in contrast to a
uniform distribution of particles expected in a gas-like system,
this liquid behavior means that the interaction in the medium of
these copiously produced particles is rather strong, as one
expects from a fluid. Detailed investigation of these phenomena
suggests that this liquid flows with almost no
viscosity~\cite{Adare:2006ti}.

\subsection{Relativistic perfect fluids}

Perfect hydrodynamics is based only on local conservation of
charge and energy-momentum and on the assumption of local thermal
equilibrium, and this is the tool that we use to describe and
calculate the properties of the matter created in relativistic
heavy ion collisions at RHIC. While there are accelerating
non-relativistic solutions in the literature, recent development
shows also relativistic solutions that can be compared to the
data~\cite{Csorgo:2006ax,Pratt:2006jj}.

Local conservation of charge and four-momentum reads as
\begin{eqnarray}
\partial_{\nu} (nu^{\nu}) & = & 0,\\
\partial_{\nu}T^{\mu\nu} & = & 0,\\
T^{\mu\nu}&=&(\varepsilon + p) u^{\mu}u^{\nu}-pg^{\mu\nu}.
\end{eqnarray}

\begin{figure}
\begin{center}
\includegraphics[height=0.6\linewidth,angle=-90]{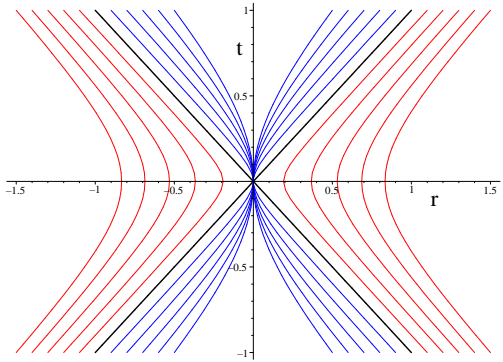}
\end{center}
\caption{\label{f:trajekt} Fluid trajectories of the new exact
solution of perfect fluid hydro,  corresponding to $d=1$ and
$\lambda = 2$. The trajectories are shown both inside and outside
the lightcone.}
\end{figure}

We find the following solution (for arbitrary $\lambda$ in $d=1$,
$\kappa=1$ and for $\lambda=2$ in arbitrary $d$ with
$\kappa=d$)~\cite{Csorgo:2006ax}:
\begin{eqnarray}
v & = & \tanh\,\lambda\eta\label{e:sol1}\\
n & = & n_0\left(\frac{\tau_0}{\tau}\right)^{\lambda d}\nu(s)\label{e:sol2},\\
T & = & T_0\left(\frac{\tau_0}{\tau}\right)^{\lambda d/\kappa
}\frac{1}{\nu(s)}\label{e:sol3},
\end{eqnarray}
where $\nu(s)$ is an arbitrary function of the scale variable $s$,
and $\eta$ is the pseudo-rapidity. As an illustration, fluid
trajectories of this solution are shown on fig.~\ref{f:trajekt}.
See details in ref.~\cite{Csorgo:2006ax}.

\section{Results}

\subsection{An advanced estimate on the initial energy density}

Based on the above solution of eqs.~\ref{e:sol1}-\ref{e:sol3} let
us estimate the initial energy density of relativistic heavy ion
or p+p reactions. As our solution is an accelerating one, and we
do not neglect the initial acceleration period, we improve the
renowned Bjorken estimate both quantitatively and qualitatively.
Similarly to Bjorken's method~\cite{Bjorken:1982qr}, we can
estimate the initial energy density (see ref.~\cite{Csorgo:2006ax}
for details). Finally we get a correction to the widely used
Bjorken formula, depending on the acceleration parameter
($\lambda$):
\begin{equation}\label{e:ncscs}
\frac{\varepsilon_0}{\varepsilon_{Bj}}=
\frac{\alpha}{\alpha-2}\left(\frac{\tau_f}{\tau_0}\right)^{1/(\alpha-2)}
\,=\,(2\lambda-1)\left(\frac{\tau_f}{\tau_0}\right)^{\lambda-1},
\end{equation}
The acceleration parameter can be extracted from the measured
rapidity distribution~\cite{Csorgo:2006ax}. For flat rapidity
distributions, $\alpha\rightarrow\infty$ ($\lambda\rightarrow 1$,
i.e.\ no acceleration) and the Bjorken estimate is recovered. For
$\lambda
> 1$, the correction factor is bigger than 1. Hence we conclude
that the initial energy densities are under-estimated by the
Bjorken formula. For realistic RHIC data from
BRAHMS~\cite{Bearden:2001qq}, the correction factor can be as big
as $\varepsilon/\varepsilon_{Bj} \approx
2.2$~\cite{Csorgo:2006ax}. Thus smaller initial bombarding (or
colliding) energies are needed to reach the critical energy
density in high energy heavy ion collisions, than thought
previously using Bjorken's renowned formula.

\begin{figure}
\begin{center}
\includegraphics[width=0.47\linewidth]{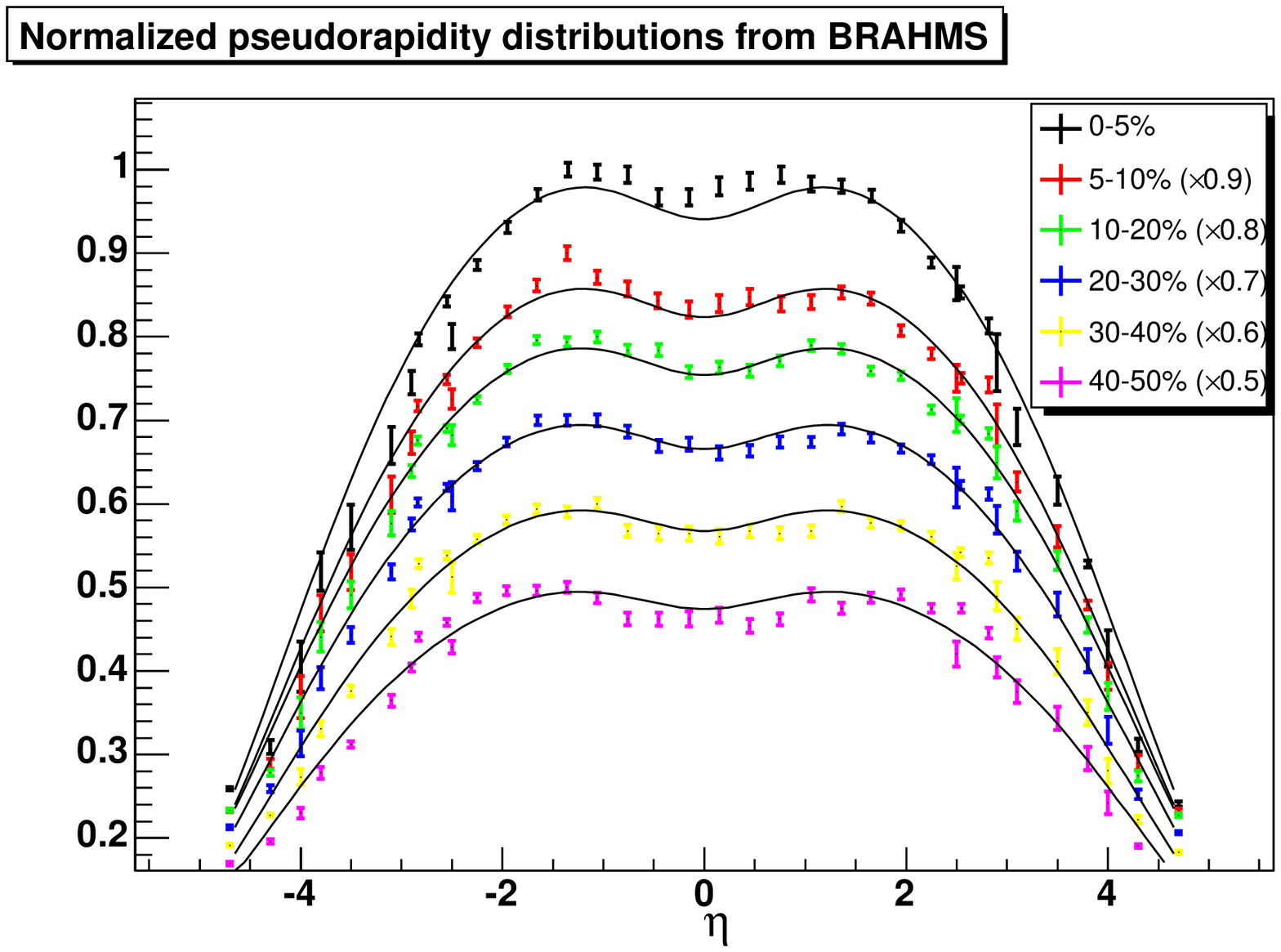}
\includegraphics[width=0.47\linewidth]{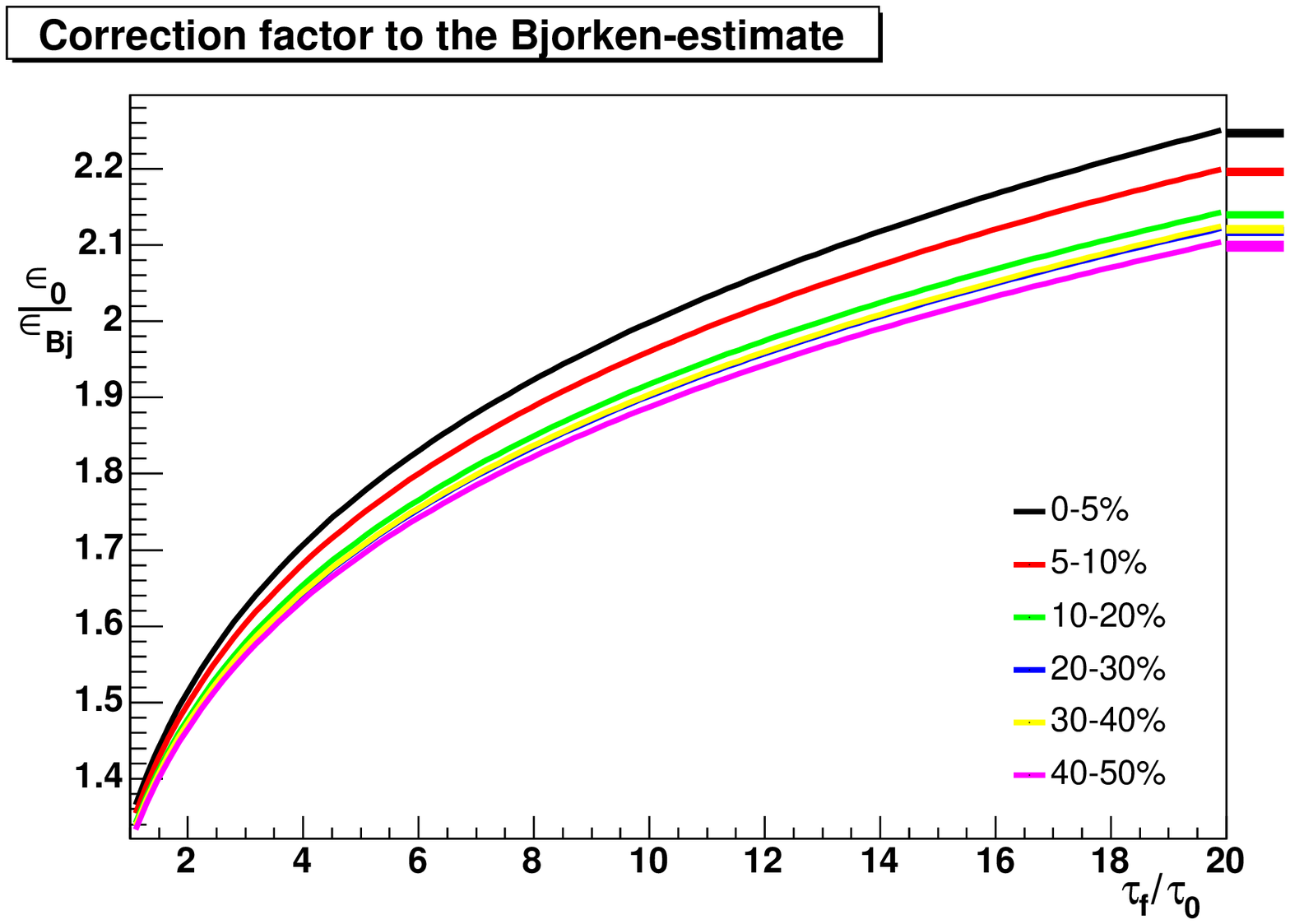}
\end{center}
\caption{\label{f:bjorkencorr} Left: charged particle
$\frac{dn}{d\eta}$ distributions of ref.~\cite{Bearden:2001qq}
fitted with the result of the relativistic hydro solution of
ref.~\cite{Csorgo:2006ax}. Right: the correction parameter
(obtained from fits show on the left panel) as a function of
freeze-out time versus thermalization time ($\tau_f/\tau_0$). At
reasonable values of 10-15, the correction to the Bjorken estimate
of energy density is a factor of $\sim$2.}
\end{figure}

\subsection{Estimating the freeze-out temperature}

\begin{figure}
\begin{center}
\includegraphics[width=0.47\linewidth]{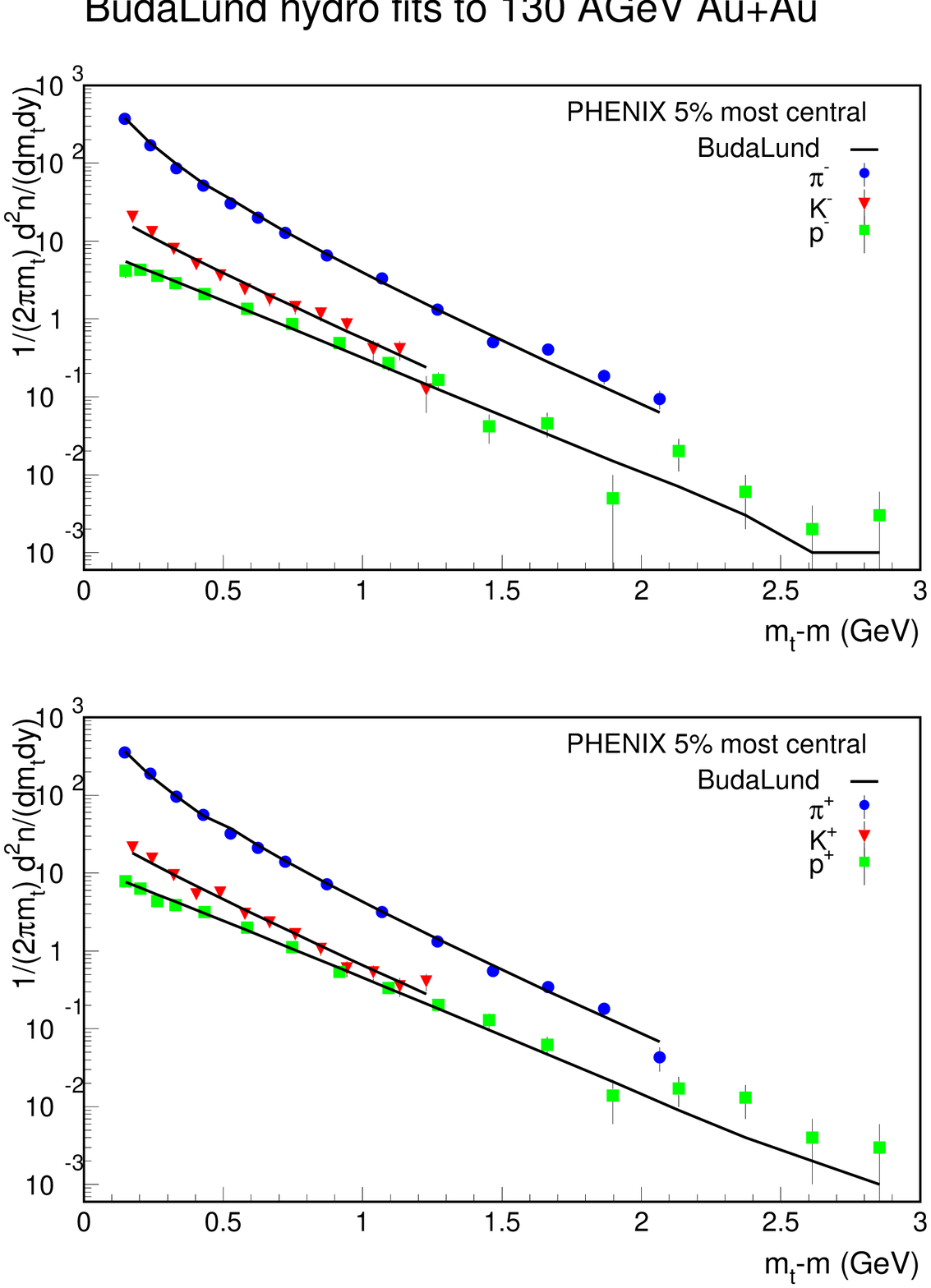}
\includegraphics[width=0.47\linewidth]{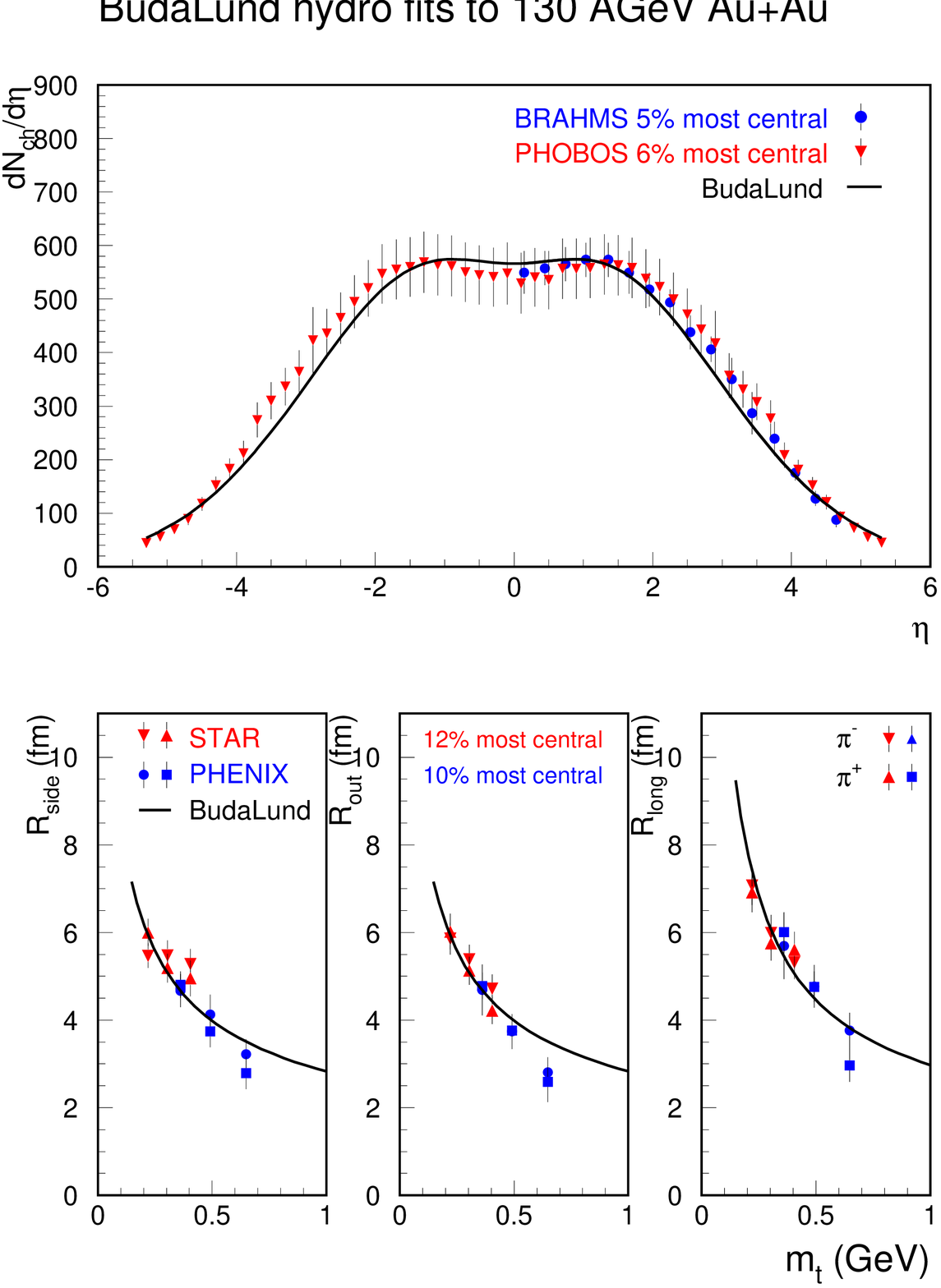} \\
\includegraphics[width=0.47\linewidth]{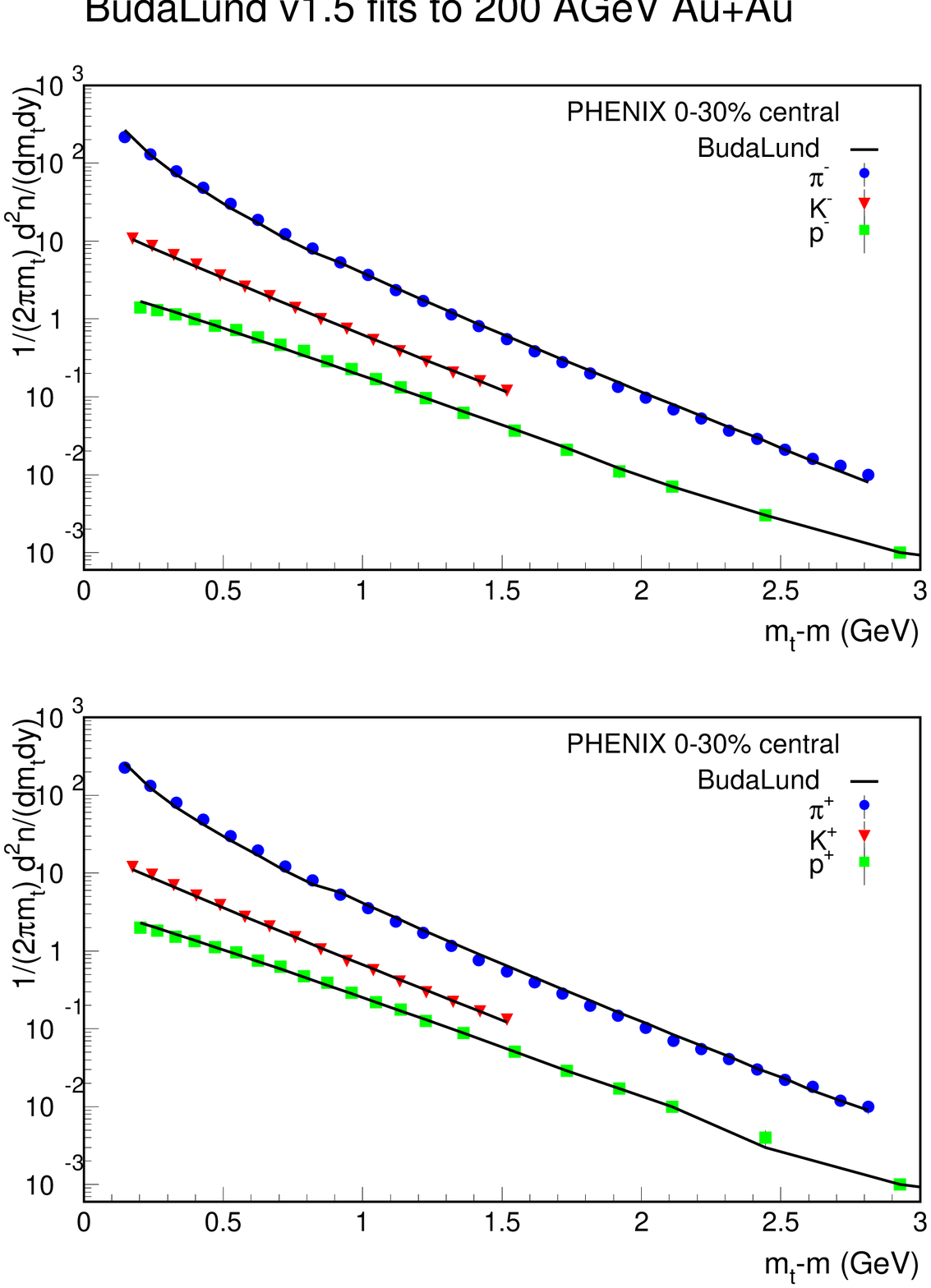}
\includegraphics[width=0.47\linewidth]{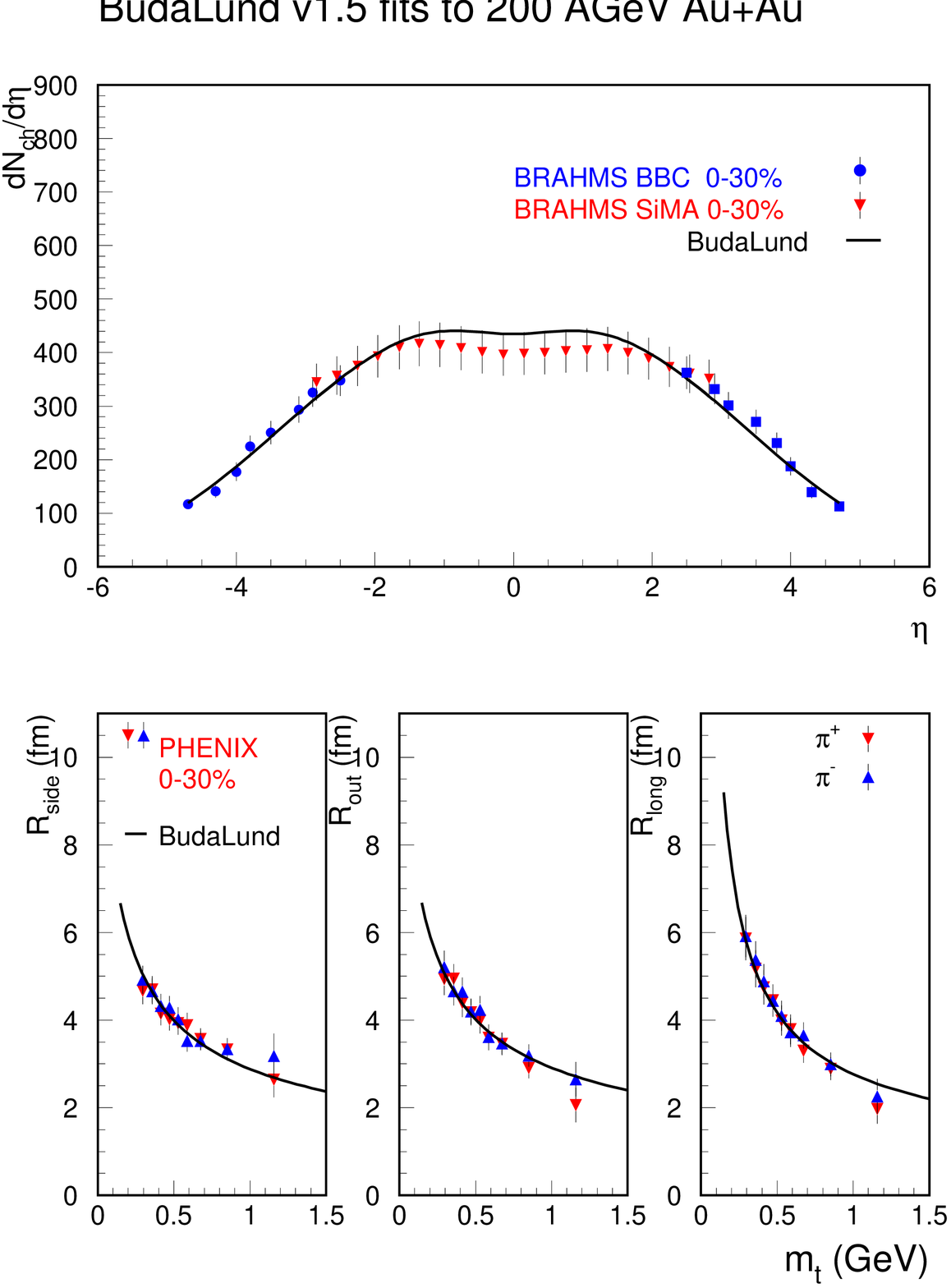}
\end{center}
\caption{\label{f:blplots}The upper four panels show a
simultaneous Buda-Lund fit  to 0-5(6)~\% central Au+Au data on
$p_t$ and $\eta$ spectra and HBT radii at $\sqrt{s_{\rm NN}} =
130$
GeV~\cite{Bearden:2001xw,Adcox:2001mf,Adcox:2002uc,Back:2001bq,Adler:2001zd}.
 The lower four  panels show similar fits to 0-30 \% central Au+Au data at
$\sqrt{s_{\rm NN}}~=~200$
GeV~\cite{Bearden:2001qq,Adler:2003cb,Adler:2004rq}. Fit
parameters are summarized in ref.~\cite{Csanad:2004mm}.}
\end{figure}

We estimated the freeze-out temperature of these Little Bangs,
fitting data to the Buda-Lund hydro
model~\cite{Csanad:2003qa,Csorgo:1995bi}. Recently, Fodor and Katz
calculated the phase diagram of lattice QCD at finite net baryon
density: their results indicate that the transition from confined
to deconfined matter is a cross-over with a nearly constant
critical temperature, $T_c = 175 \pm 2$ MeV~\cite{Aoki:2006br}.
The result of the Buda-Lund fits to RHIC Au+Au data of
refs.~\cite{Bearden:2001qq,Bearden:2001xw,Adcox:2001mf,Adcox:2002uc,Back:2001bq,Adler:2001zd,Adler:2003cb,Adler:2004rq}
(shown on fig.~\ref{f:blplots}), in particular the value of the
fit parameter $T_0$ (central freeze-out temperature, see details
in ref.~\cite{Csanad:2004mm}), indicates the existence of a region
several standard deviations hotter than the critical temperature.
This is an indication on quark deconfinement in Au + Au collisions
with $\sqrt{s_{\rm NN}} = 130 $ and $200 $ GeV at
RHIC~\cite{Csanad:2004mm,Csanad:2003sz,Csanad:2004cj}, confirmed
by the analysis of $p_t$ and $\eta$ dependence of the elliptic
flow~\cite{Csanad:2003qa}. A similar analysis of Pb+Pb collisions
at CERN SPS energies yields central temperatures lower than the
critical value, $T_0<T_c$~\cite{Ster:1999ib,Csorgo:1999sj}.

\subsection{Universal scaling of the elliptic flow}

The Buda-Lund calculation of the elliptic flow results (under
certain conditions detailed in ref.~\cite{Csanad:2003qa}) in the
following simple universal scaling law:
\begin{equation}
v_2=\frac{I_1(w)}{I_0(w)},\label{e:v2w}
\end{equation}
thus the model predicts a \emph{universal scaling:} every $v_2$
measurement is predicted to fall on the same \emph{universal}
scaling curve $I_1/I_0$ when plotted against the scaling variable
$w$ (see details in ref.~\cite{Csanad:2005gv}).

This means, that $v_2$ depends on any physical parameter
(transverse or longitudinal momentum, center of mass energy,
centrality, type of the colliding nucleus etc.) only through the
(universal) scaling parameter $w$.

\begin{figure}
\begin{center}
  \includegraphics[width=1\linewidth]{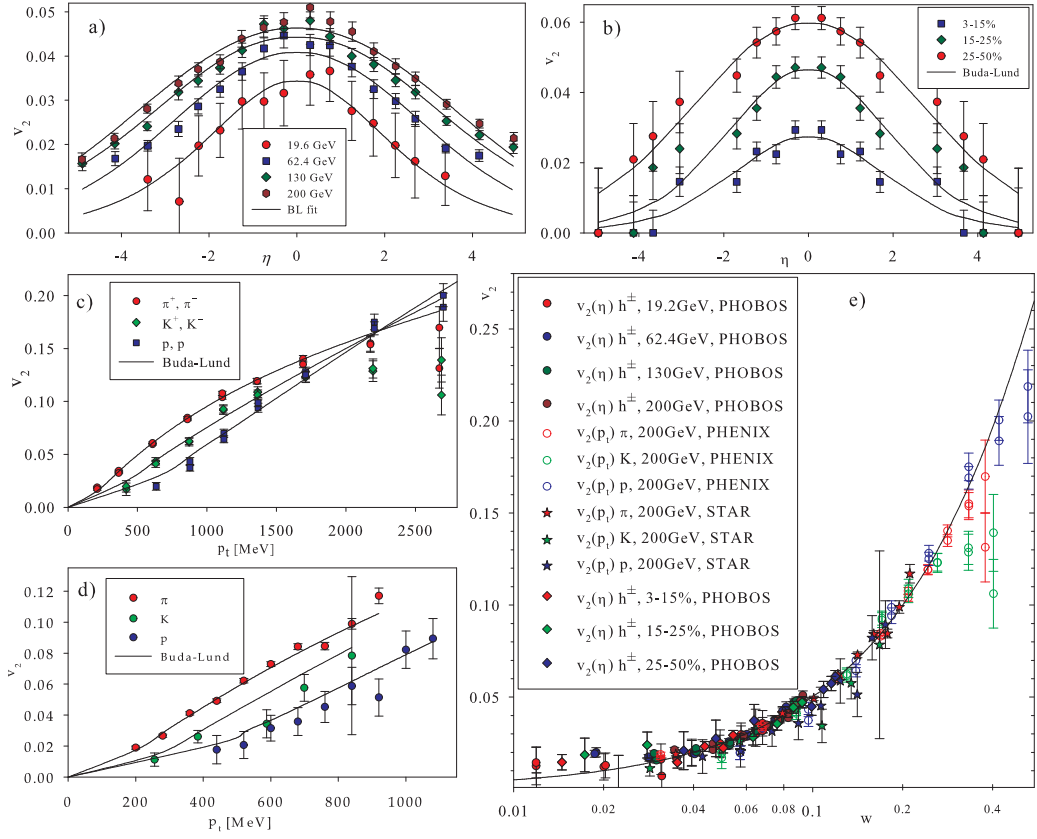}\\
\end{center}
\caption{PHOBOS~\cite{Back:2004zg} (a-b),
PHENIX~\cite{Adler:2003kt} (c) and STAR~\cite{Adams:2004bi} (d)
data on elliptic flow, $v_2$, plotted versus $p_t$ and $\eta$ and
fitted with Buda-Lund model. Elliptic flow versus variable $w$ is
plotted in panel (e): data points of plots (a-d) show the
predicted~\cite{Csanad:2003qa} universal scaling. See fit
parameters in ref.~\cite{Csanad:2005gv}} \label{f:v2fit}
\end{figure}

In ref.~\cite{Csanad:2005gv} we have shown that the excitation
function of the transverse momentum and pseudorapidity dependence
of the elliptic flow in Au+Au collisions (RHIC data from
refs.~\cite{Adler:2003kt,Back:2004zg,Adams:2004bi}) is well
described with the formulas that are predicted by the Buda-Lund
type of hydrodynamical calculations. We have provided a
quantitative evidence of the validity of the perfect fluid picture
of soft particle production in Au+Au collisions at RHIC up to
~1-1.5\,GeV but also show here that this perfect fluid extends far
away from mid-rapidity, up to a pseudorapidity of
$\eta_{\textrm{beam}}-0.5$. The universal scaling of PHOBOS
$v_2(\eta)$~\cite{Back:2004zg}, PHENIX
$v_2(p_t)$~\cite{Adler:2003kt} and STAR
$v_2(p_t)$~\cite{Adams:2004bi}, expressed by Eq.~(\ref{e:v2w}) and
illustrated by Fig.~\ref{f:v2fit}.e provides a successful
quantitative as well as qualitative test for the appearance of a
perfect fluid in Au+Au collisions at various colliding energies at
RHIC.

\subsection{Chiral symmetry restoration}
Correlation functions are important to see the collective
properties of particles and the space-time structure of the
emitting source, e.g.\ the observed size of a system can be
measured by two-particle Bose-Einstein
correlations~\cite{HanburyBrown:1956pf}. The $m_t$ dependent
strength of two-pion correlations, the so-called $\lambda_*$
parameter, which is related to the extrapolated value of the
correlation function at zero relative momentum, can be used to
extract information on the mass-reduction of the $\eta$' meson, a
signal of $U_A(1)$ symmetry restoration in the
source~\cite{Vance:1998wd,Kapusta:1995ww,Huang:1995fc,Hatsuda:1994pi}.

\begin{figure}[t]
   \begin{center}
   \includegraphics[width=0.32\linewidth]{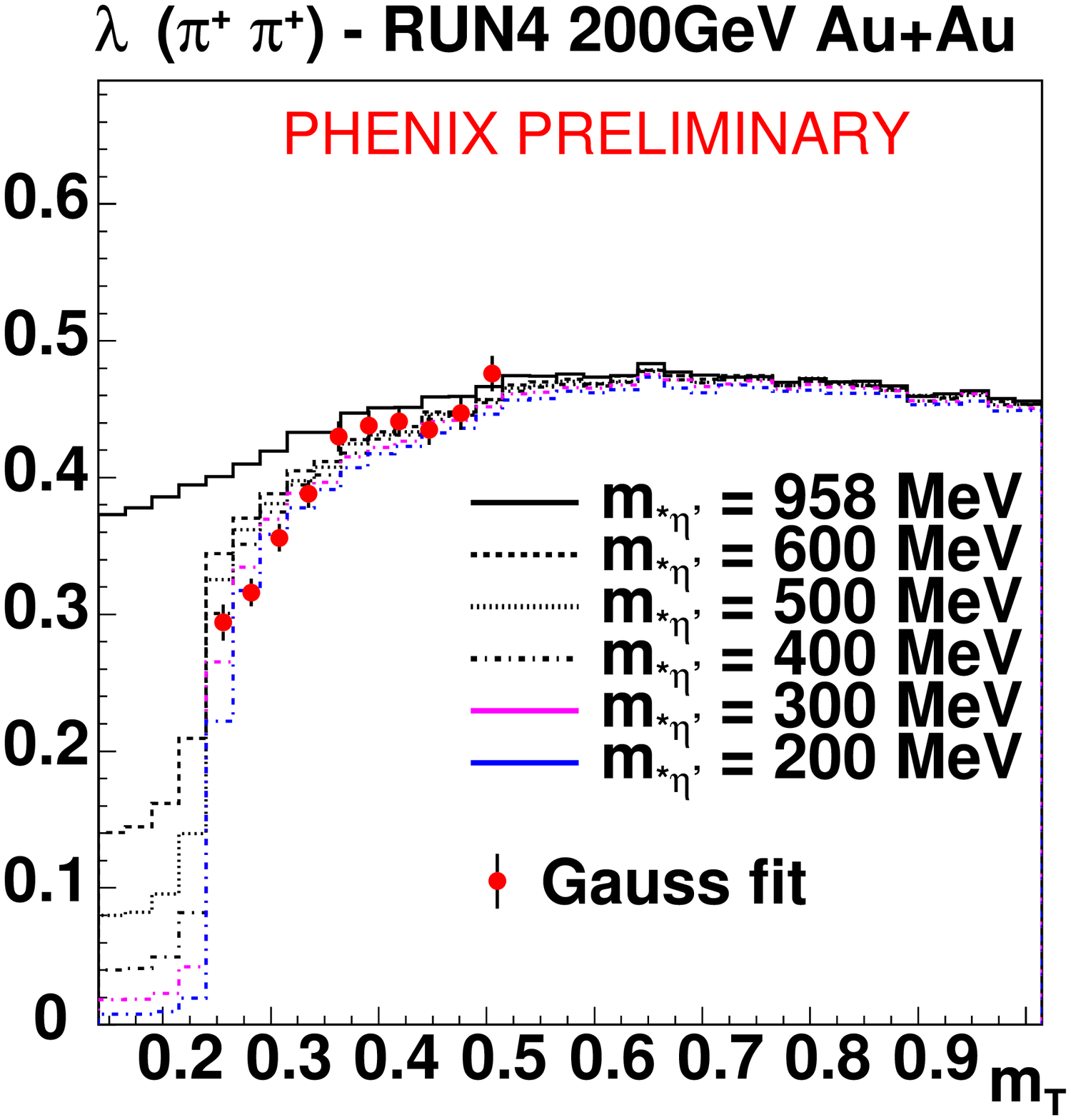}
   \includegraphics[width=0.32\linewidth]{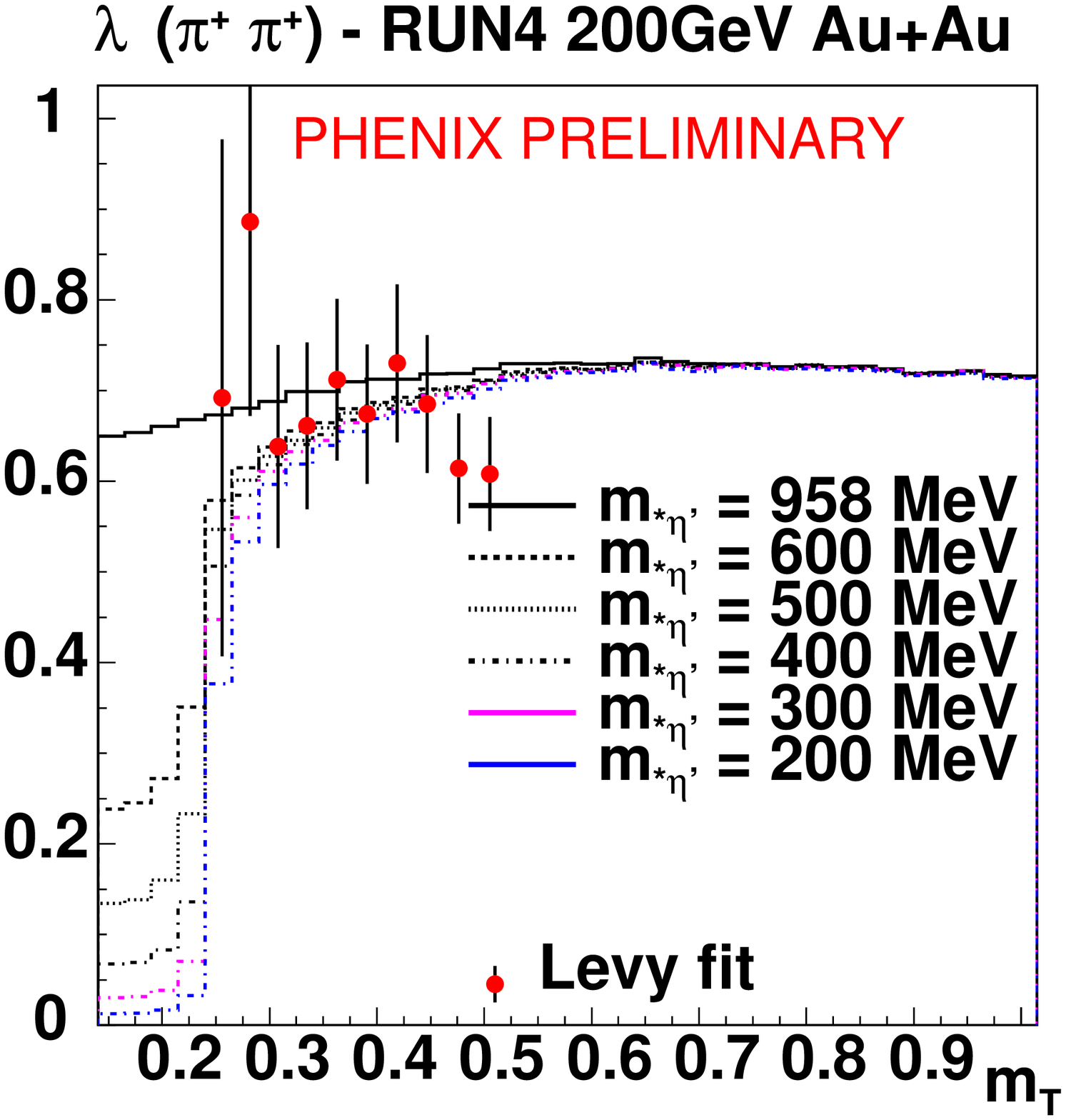}
   \includegraphics[width=0.32\linewidth]{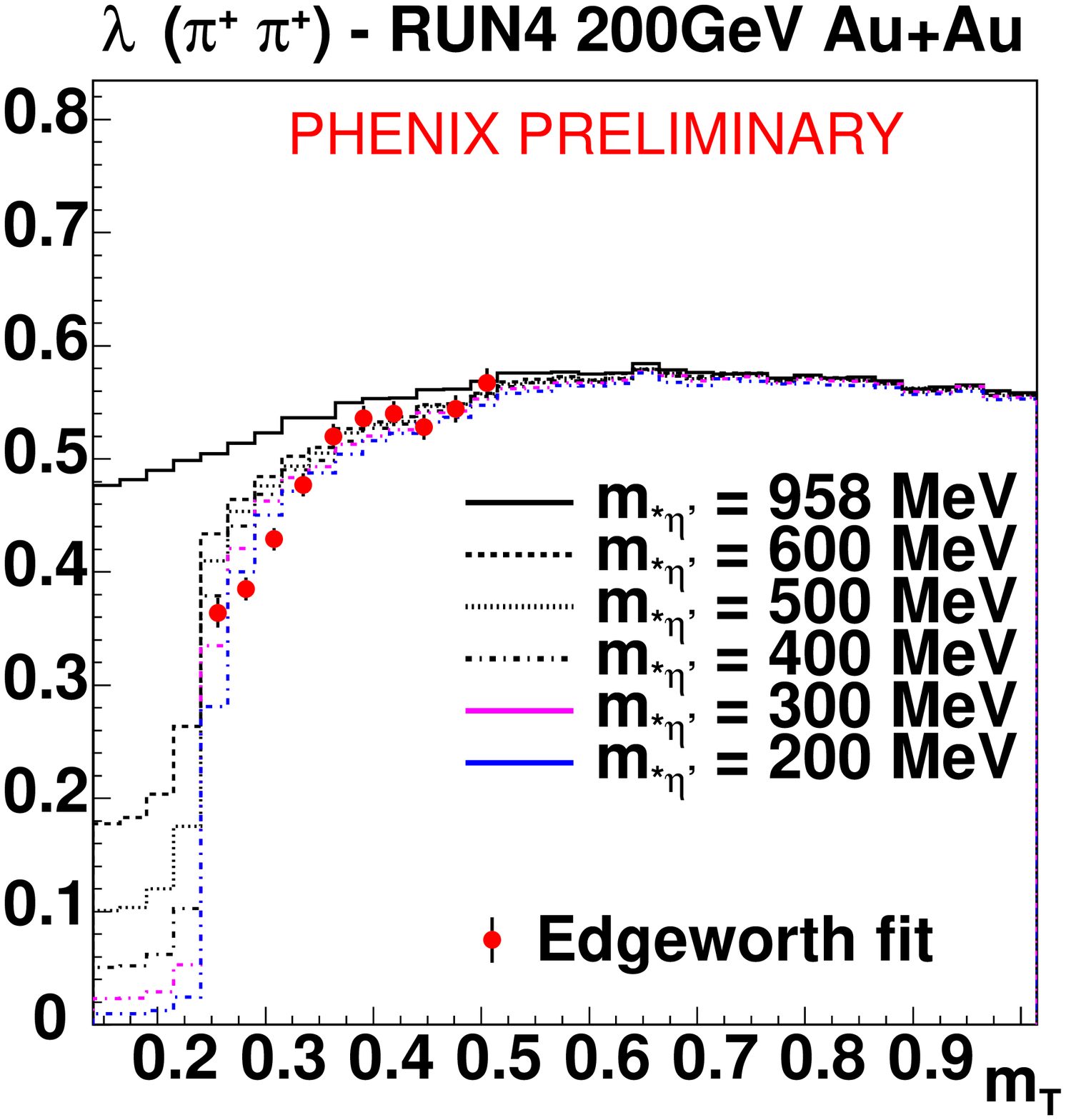}\\
   \includegraphics[width=0.9\linewidth]{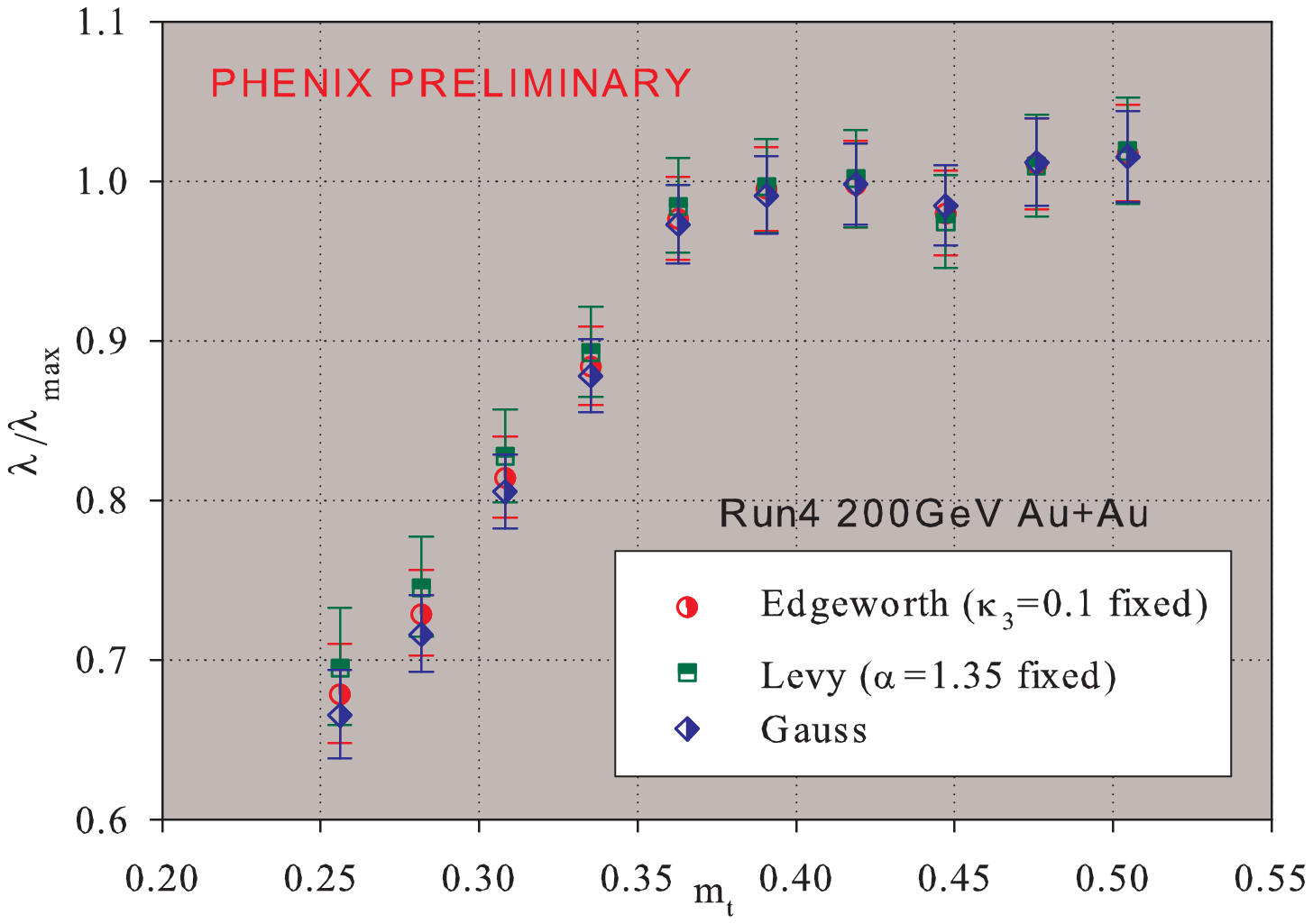}
   \end{center}
\caption{Top three figures: measured $\lambda_*(m_t)$ of
ref.~\cite{Csanad:2005nr} compared to calculations using the model
of ref.~\cite{Vance:1998wd} with various $\eta'$ mass values.
Bottom figure: $\lambda_*(m_t)$ curves with equal number of fit
parameters and re-normed with their maximal value on the interval
of 0.20GeV$<m_t<$0.55GeV all show the same shape. See details of
the fits in ref.~\cite{Csanad:2005nr}.}\label{f:ua1}
\end{figure}

PHENIX analyzed~\cite{Csanad:2005nr} $\lambda_*(m_t)$ with fits to
two-pion correlation functions using three different shapes,
Gauss, Levy and Edgeworth, described in
refs.~\cite{Csorgo:1999sj,Csanad:2005nr,Csorgo:2003uv}. A
comparison of the measurements with model calculations of
ref.~\cite{Vance:1998wd} using FRITIOF results for the composition
of the long-lived resonances and a variation of the $\eta'$ mass
is presented in fig.~\ref{f:ua1}. If we re-norm the
$\lambda_*(m_t)$ curves with their maximal value on the
investigated $m_t$ interval, they overlap, confirming the
existence and characteristics of the hole in the $\lambda_*(m_t)$
distribution.

Gauss fit results agree with former PHENIX measurements (see
ref.~\cite{Adler:2004rq}). Regarding $U_A(1)$ symmetry
restoration, conclusion is that at present, results are critically
dependent on the understanding of statistical and systematic
errors, and additional analysis is required to make a definitive
statement.

\section{Summary and conclusions}

In summary, we can make the definitive statement, based on
elliptic flow measurements and the broad range success of analytic
hydro models, that in relativistic Au+Au collisions observed at
RHIC we see a perfect fluid. Based on our estimates on the
temperature and energy density we also conclude that the observed
matter is in a deconfined state. We also see a possible signal of
partial symmetry restoration in the mass reduction of $\eta$'
bosons. Future plan is to explore all properties of the Quark
Matter, by analyzing more data and using higher luminosity. We are
after the full map of the QCD phase diagram, and in order to
explore it, we also have to go to higher energies and compare them
to lower energy data. If the Quark Matter is the New World, then
Columbus just realized he is not in India, but on a new continent.

\begin{quote}
{\small {\it ``It does not make any difference how beautiful your
guess is. It does not make any difference how smart you are, who
made the guess, or  what  his name is --- if it disagrees with
experiment it is wrong.''}}

R. P. Feynman, about discovering new laws~\cite{Feynman}
\end{quote}

%

\bibliographystyle{prlunsrt}
\bibliography{Master}

\end{document}